\documentclass[12pt]{article}
\usepackage{amssymb}
\usepackage{amscd}

 \linethickness{0.4pt}

\begin{document}
\newtheorem{lemma}{Lemma}[section]
\newtheorem{theorem}{Theorem}[section]
\newtheorem{corollary}{Corollary}[section]
\newtheorem{definition}{Definition}[section]
\begin{flushright}
                IFT UwB 01/2006 \\
\end{flushright}

\bigskip
\bigskip
\begin{center}
\section*{{
Anomalus BRST Complexes for Non-Critical Massive Strings.}}
\end{center}
\bigskip
\bigskip
\begin{center}
{{\large\bf Marcin Daszkiewicz}}
\end{center}
\begin{center}
{
{Institute of Theoretical Physics\\ Wroc{\l}aw University  pl.Maxa
Borna 9, 50-206 Wroc{\l}aw, Poland\\ E-mail:
marcin@ift.uni.wroc.pl}}
\end{center}
\begin{center}
{\large\bf Zbigniew Hasiewicz, Cezary J. Walczyk }
\end{center}
\begin{center}
{
{Institute of Theoretical Physics\\ University in Bia{\l}ystok,
ul. Lipowa 41, 15-424 Bia{\l}ystok, Poland\\ E-mail:
zhas@physics.uwb.edu.pl,\\~~~~~~~~~~~~~c.walczyk@alpha.uwb.edu.pl}}
\end{center}
\bigskip\bigskip

\date{November 2005}

\begin{abstract}
\noindent {It is shown that the BRST resolution of the spaces of
physical states of non-critical (anomalus) massive string models
can be consistently defined. The appropriate anomalus complexes
are obtained by canonical restrictions of the ghost extended
spaces to the kernel of curvature operator and additional
Gupta-Bleuler like conditions without any modifications of the
matter sector. The cohomologies of the polarized anomalus complex
are calculated and analyzed in details.}
\end{abstract}
\vspace{1.5cm}
\thispagestyle{empty}
\eject
\section*{Introduction}

The massive string model was introduced in its full extent in the
paper \cite{hasjas} as a step towards the consistent and effective
description of the strings (treated as 1-dimensional extended
objects) in sub-critical dimensions $1<d<26$. The classical model
was defined by the variational principle with standard Nambu-Goto
action functional (in its BDHP form \cite{bdhp}) supplemented by
the term to govern the motion of an additional scalar - the
Liouville field.\\ On the level of canonical formulation, the
massive string model is the constrained system of mixed type.
Almost all constraints (except of the kinetic one) are of second
class.

The standard quantization procedure applied to the classical
massive string results in relativistic, constrained quantum system
with anomaly. The anomaly contains both classical and quantum
contributions. The presence of the anomaly enforces the
application of the Gupta-Bleuler like procedure \cite{gb}
(polarization of constraints) to define the subspace of physical
degrees of freedom (\ref{phys}).\\
    By the use of the general results on the structure
of the Verma modules of the Virasoro algebra \cite{noghost} it was
proved that the family of unitary (and formally also
relativistically invariant) quantum massive string models can be
distinguished. This family decays into two series: continuous
(\ref{cont_series}) and discrete one (\ref{disc_series}).\\
     The particle content of these series was
    analyzed and completely determined in \cite{spectrum}.
    The form of their spectra makes them interesting and
    promising from the point of view of the applications to the
    description of the interactions of low-energy (composite) QCD
    states.\\
    \\
    The interactions of  one-dimensional extended objects
    appeared to be most efficiently described within the framework
    of cohomological BRST formalism \cite{brstinteractions}. On
    the other hand it was shown \cite{brstn} that only critical massive
    strings (\ref{critical_values}) located at the very beginning of the discrete
    series  do admit the BRST resolution in its standard form. The
    reason is that only for these critical models the curvature operator (the square of differential)
    does vanish identically - the ghost-extended system is anomaly
    free.\\
    \\
    This paper is devoted to the natural generalization of the
    canonical BRST approach, which enables one to look at all
    models from the universal point of view. The formalism
    can be applied to the whole class of unitary
    strings\footnote{From the formal point of view it can be applied to all non-unitary models too.}
    containing both: critical (anomaly free) and non-critical (anomalus) models.
    The proposed approach, in contrast to the schemes applied
    earlier \cite{fradkin},\cite{GB} does not introduce any additional
    degrees of freedom -  neither in matter nor in ghost sector.
    The
    main idea consists in restricting the appropriate differential
    space. \\
    There are two complexes which one may associate with anomalus models.
    The first one, called the anomalus complex, is defined as the subspace
    annihilated by the anomaly (curvature)
    operator - the square of the canonically constructed BRST
    differential. The cohomologies of this complex are
    non-vanishing at higher ghost numbers and are quite
    complicated to be completely determined. \\
    The second complex, called the polarized anomalus
    complex, provides the cohomological resolutions of the
    spaces of physical states for all unitary models. This means that
    the critical model can be thought of as the limiting case also
    from the point of view of BRST formulation - in this sense that the underlying
    space of the corresponding complex is the largest possible.\\
    It is worth to make one comment on the problem of null vectors in the
    space of physical states (\ref{phys}) in the context of BRST
    formulation. In the critical case all the null states (radical) are
    cohomologically trivial and can be eliminated by BRST-gauge
    transformations \cite{brstn}. The corresponding No-Ghost
    Theorem can be proved by very general cohomological methods
    \cite{brstinteractions},\cite{Mordechai}. \\
    The results presented in this paper indicate that the relation
     between the presence of null states and
    (cohomological) gauge symmetries can be directly established
    in the critical  case only. Indeed, the  Theorem
    $3.1$ clearly states that there are non-trivial relative cohomology
    classes, which according to e.g. \cite{difrancesco} are null.
    A better understanding of this appearence needs further
    investigation.\\
    \\
    The considerations of this paper are restricted to massive string
    models for at least two reasons. On the one hand side the anomalies present in these models
    are of mixed nature: there are classical and quantum contributions. On the other hand
    their structure is simple enough to perform  the detailed analysis of polarized anomalus
    cohomologies.\\
    It is however evident that the proposed approach can be
    applied to  other constrained systems too. It should give the
    analogous results, namely cohomological resolutions of the
    corresponding spaces of physical states by the polarized complex. It should also be
    mentioned that the property that the elements of the anomalus complex
    are the highest weight vectors of  $sl(2,\mathbb{C})$ algebra (\ref{sldwa}),(\ref{com}) generated by
    the anomaly operator seems to be universal within the wide class of constrained
    systems.\\
    \\
      The paper is organized as follows. \\
      In the first section, for the
sake of completness, the sketch of massive string model is briefly
presented. The No-Ghost Theorem is recalled in order to restrict
further considerations to unitary non-critical massive string
models. \\ In the second chapter the anomalus complexes for string
models are constructed: they are defined as subspaces  of standard
ghost extended string spaces (called string differential spaces
throughout this paper). It appears that anomalus complexes inherit
the essential properties of critical massive string BRST complex,
in particular the bigraded structure of the relative  cochains.\\
    The next chapter is devoted to the computation and analysis of
 the cohomologies of the polarized anomalus BRST complex. The bigraded
cohomologies (Dolbeaut-Grotendieck) as well as  relative
cohomologies are calculated and represented in terms of physical
string states of ~"old covariant formalism". It is thus  shown,
that also in non-critical case, the polarized anomalus BRST
complex provides the resolution of the space of physical string
states. The absolute cohomologies of the polarized complex are
also reconstructed.
\\ Finally, the concluding remarks are added, and some open
question are raised.
    \eject

\section{The classical and quantum massive strings}
A brief description of the classical and quantum massive string
models \cite{hasjas} is presented in this section.


\subsection{The classical massive string}
The classical massive string model is defined by the functional
\cite{hasjas}:
\begin{eqnarray}
S[\,M,\;g,\;\varphi,\;x\,] = &-&\frac{\alpha}{2\pi}\int_M \sqrt{-g}
\;d^2z\;g^{ab}\partial_ax^\mu \partial_bx^\nu \cr &~~& \cr
&-&\frac{\beta}{2\pi}\int_M \sqrt{-g}
\;d^2z\left(g^{ab}\partial_a\varphi
\partial_b\varphi +2R_g\varphi \right ) \;\;,\label{action}
\end{eqnarray}
which is an extension of the standard $d$-dimensional string
world-sheet action by an additional term, which governs the motion
of the scalar $\varphi(z)$ called the Liouville field. \\
    The detailed analysis \cite{hasjas}
of the variational problem for (\ref{action}) leads to the
constrained phase space system. The phase space of this system is
parametrized by real canonical pairs $(x^{\mu},p_{\mu})$ with
\begin{eqnarray}
    \label{classzero}
    \{\,x^{\mu},p_{\nu}\,\} = \delta^{\mu}_\nu\;,\;\;\;\mu,\nu = 0 , \ldots, d - 1 \;\;,
\end{eqnarray}
 to describe the center-of-mass motion, and the complex
variables $a^{\mu}_{m},\;u_{m}$ $(m \in \mathbb{Z} \setminus \{0\})$
related to the oscillatory degrees of freedom. The variables
$a^{\mu}_{m},\;u_{m}$ are constructed out of higher Fourier modes of
the real fields $x^{\mu}(z)$ and $\varphi(z)$ and their canonically
conjugated momenta. Their Poisson brackets are  of standard form:
\begin{eqnarray}
    \label{classccr}
    \{\,a^{\mu}_{m},a^{\nu}_{n}\,\} = im\eta^{\mu\nu}\delta_{m+n}\;,
    \;\; \{\,u_{m},u_{n}\,\} = im\delta_{m+n}\;,
    \;\;\mu,\nu = 0 , \ldots, d - 1 \;\;.
\end{eqnarray}

    The classical constraints of the system correspond to, roughly speaking,
    the invariance of (\ref{action}) with respect to the diffeomorphisms of the world-sheet.
    They are given by the generators of  hamiltonian action of the gauge group on the phase space.
    The complex modes of the canonical generators of this action are given by the following expressions:
\begin{eqnarray}
    l_{m}\,= \frac{1}{2}\, \sum_{{k\in\mathbb{Z}}}\,a_{m+k}\cdot a_{- k}
    \, +  \frac{1}{2}\, \sum_{k\in {\mathbb{Z}}}\,u_{m+k}u_{- k}
     +2i\sqrt{\beta}\,m\,u_{m}\;,\;\; m \in \mathbb{Z}\;.
     \label{classconstraints}
\end{eqnarray}
    The Poisson bracket algebra of the above modes does not close.
    It is modified by non-zero central term - the Lie algebra scalar cocycle:

\begin{equation}
    \left \{\,l_{m} , l_{n}\,\right \} = i( m - n ) l_{m + n} - 4i\beta m^3\delta_{m+n}
    \;.
    \label{classvirasoro}
\end{equation}
The presence of the central term in the Poisson bracket relations
(\ref{classvirasoro}) indicates that the constraints are of mixed
type with $l_0$ being the unique constraint of first class. The
corresponding gauge group action is called \cite{libermann} weakly
hamiltonian in this case.


\subsection{The quantum massive string - the states in "old covariant quantization"}
{\bf Basic definitions} \\
\\
The space of states of massive string model is most conveniently
defined as a direct integral:
\begin{equation}
 {\cal H}\; = \;\int d^{d}p\;{\cal H}(p)\;\;,  \label{covariant space}
\end{equation}
of pseudo-unitary Fock spaces. Every ${\cal H}(p) $ is generated by
the algebra of excitation operators:
\begin{eqnarray}
[\,a^{\mu}_{m},a^{\nu}_{n}\,] &=&
m\eta^{\mu\nu}\delta_{m+n}\;\;,\cr [\,u_{m},u_{n}\,] &=&
m\delta_{m+n}\;\;\;\;\;\;\;\;\; m,n \in {\mathbb{Z}}\;\;\;\mu,\nu
= 0 , \ldots, d - 1 \;\;. \label{ccr}
\end{eqnarray}
out of the vacuum vector $\omega(p)$ satisfying $a^{\mu}_{m}\omega
(p) = 0 = u_{m}\omega(p)\;;\;m > 0$. The vacuum vectors are
generalized eigenfunctions for the momentum operators:
$P^{\mu}\omega(p) = p^{\mu}\omega(p)$ and are formally normalized by
the condition: $\left(\omega(p),\omega(p'\right))= \delta(p-p')$. \\
The unique scalar product in (\ref{covariant space}) is defined by
imposing formal conjugation rules $a^{\mu
*}_{m} = a^{\mu}_{- m},\;u_{m}^{*} = u_{- m}$ on string modes. \\
The operators corresponding to the constraints are obtained by
replacing the classical modes (\ref{classccr}) by their operator
counterparts in the expressions (\ref{classconstraints}) and
normal ordering:
\begin{eqnarray}
L_{m}\,= \frac{1}{2}\, \sum_{{k\in\mathbb{Z}}}\,:a_{m+k}\cdot a_{-
k}: \, + \, \frac{1}{2}\, \sum_{k\in {\mathbb{Z}}}\,:u_{m+k}u_{- k}:
 +\,2i\sqrt{\beta}\,m\,u_{m} \,+\,2\beta \delta_{m
0}\;\;. \label{virasoro}
\end{eqnarray}
The classical central term in (\ref{classvirasoro}) gets modified
by normal ordering anomaly:
\begin{equation}
\left[\,L_{m} , L_{n}\,\right] = ( m - n ) L_{m + n} +
\frac{1}{12} m\left(m^2 - 1\right) \left(d + 1 + 48 \beta \right)
\delta_{m + n}\;\;. \label{virasorocom}
\end{equation}
The anomaly in the quantum constraints algebra has two sources.
There is trace of the classical central term and strictly quantum
contribution.\\
\\
{\bf The space of physical states and No-Ghost Theorem}\\
\\
    Within the framework of ~"old covariant"
quantization procedure, the space of physical states  is defined
as the subspace of (\ref{covariant space}) consisting of vectors
subject to  the infinite set of equations:
\begin{equation}
{\cal{H}}_{\rm phys} = \left\{\,\Psi \;\;;\;\; \left( L_{ m } - a
\delta_{m 0} \right)\Psi  =  0 \;,m\geq 0\,\right\}\;\;.
    \label{phys}
\end{equation}
The real parameter $a$ defines the beginning of the string mass
spectrum and is left to be fixed by natural consistency
conditions: the unitarity  and relativistic invariance. The
unitarity condition means that the inner product in the space
(\ref{phys}) should be non-negative. Then the quotient of
(\ref{phys}) by the subspace of all null vectors (radical) gives
(after completion) the Hilbert space of states.


According to the {No-Ghost} theorem \cite{hasjas} one can
distinguish two principal families of unitary string models.
 \\
 There is continuous series:
\begin{equation}
\label{cont_series}
 a \leq 1 \;\;\;,\;\;  0< \beta \leq {24-d \over 48}\;\;,
\end{equation}
and discrete one:
\begin{eqnarray}
\label{disc_series}
    \beta & = & {24-d \over 48}+{1\over 8m(m+1)}
\; ; \;\;\;\;\;\; \;\; m\geq 2  \cr &~~& \cr
a & = &
1-{((m+1)r-ms)^2 -1 \over 4m(m+1)} \; ; \;\; 1\leq r \leq m-1 ,\;
1\leq s \leq r \;\;.
\end{eqnarray}
The beginning of the discrete series ($m=2$) corresponds to so
called critical massive string model with:
\begin{equation}
\beta = \frac{25 - d}{48}\;\;\;\;{\rm and}\;\;\;\; a = 1\;\;.
\label{critical_values}
\end{equation}
It was shown in \cite{brstn} that in this case the space of
physical states admits standard BRST resolution.  It was also
shown that in the case of (\ref{critical_values}) all the elements
of the radical correspond to pure gauge degrees of freedom in the
sense of BRST cohomology.
\\
The construction of appropriate BRST-like complexes for massive
string models corresponding to non-critical (anomalus) values of the
parameters (\ref{cont_series},\ref{disc_series}) will be given in
the next section.


\section{Anomalus BRST complexes}

 For the reader's convenience the section is divided into
three parts containing ingredients and subsequent steps of the
construction. The first two steps are common to both: critical and
anomalus models and may be found in the literature (e.g.
\cite{GHV}). They are repeated here for the sake of completeness
and in order to stress that one is able to split the differential
spaces and corresponding complexes into finite dimensional
subspaces and subcomplexes. This splitting  plays an essential
role in the computation of cohomologies.


\subsection{The canonical ghost sector}
In order to construct the anomalus complexes one starts with the
same ghost differential space\footnote{Differential space is by
definition a pair $(\,\cal{C}, D\,)$ consisting
 of graded space $\cal{C}$ and differential $D$ of order $\pm 1$,
 which is not necessarilly nilpotent.}
as in the case of critical massive string models \cite{brstn}.
Despite of the presence of the anomaly (\ref{virasorocom}) the
quantum constraints (\ref{virasoro}) are treated as if they were
of first class.

   With every constraint $L_m\;,\; m \in \mathbb{Z}$ (\ref{virasoro}) one associates a pair of fermionic operators:
   the ghost $c_{m}$ and the anti-ghost $b_{m}$. It is assumed that they satisfy
the following anti-commutation relations:
\begin{equation}
\left\{\,b_{m},c_{n}\,\right\} = \delta_{m + n}\;\;,\;\;
\left\{\,b_{m},b_{n}\,\right\} =
0\;\;,\;\;\left\{\,c_{m},c_{n}\,\right\} = 0\;\;,\;\; m,n \in
\mathbb{Z}\;\;. \label{ghosts}
\end{equation}
The ghost sector ${\cal{C}}_{\rm gh}$  is created out of the
vacuum of the ghost space, which is formally defined as a vector
$\omega $ satisfying:
\begin{equation}
c_{m} \omega = 0\;\;;\; m > 0\;\;\;\; {\rm and }\;\;\;\; b_{m}\omega
= 0 \;\;;\;m\geq 0 \;\;. \label{ghostvacuum}
\end{equation}
The space ${\cal{C}}_{\rm gh}$ is equipped with the
(non-degenerate) scalar product by imposing the following
normalization condition: $\left(\,\omega , c_{0} \omega\, \right)
= 1$ of the ghost vacuum, and the following formal conjugation
rules:
\begin{equation}
b_{m}^{*} = b_{- m}\;\;,\;\; c_{m}^* = c_{- m }\;\;, \label{con}
\end{equation}
of the ghost modes.

The differential on the space ${\cal{C}}_{\rm gh}$ is constructed
in most convenient  way \cite{GHV} with the help of the natural
realization of the constraint algebra on the ghost space:
    \begin{equation}
    \label{vir_ghost}
 {\cal{L}}_{m} = \sum_{k\in\mathbb{Z}}
(k-m):c_{- k} b_{m + k}\!: \;\;.
    \end{equation}
The normal ordering prescription in the formulae above is
determined, as usually, by the conditions  imposed on the vacuum
(\ref{ghostvacuum}) and is supplemented with the
antisymmetrization rule for the ghost zero-modes $\;:c_0b_0:\; =
\frac{1}{2}(c_0b_0 - b_0c_0)$. The resulting operators
${\cal{L}}_{m}\;,\; m \in \mathbb{Z}$
 satisfy the following Virasoro algebra relations:
 \begin{eqnarray}
 \left[\,{\cal{L}}_{m}\,,\,{\cal{L}}_{n}\,\right] =
(m-n){\cal{L}}_{m+n} - \delta_{m+n}( \frac{26}{12}(m^3 - m) + 2m
)\;\;. \label{ghostvir}
 \end{eqnarray}
 The differential in ${\cal{C}}_{\rm gh}$ is defined to be:
\begin{equation}
d_{\rm gh} = \frac{1}{2}\sum_{m>0} c_{-m}\: {\cal{L}}_{m} +
\frac{1}{2}\, c_{0}\, {\cal{L}}_{0} + \frac{1}{2} \sum_{m>0}
{\cal{L}}_{- m}\,c_m\;\;, \label{ghostdiff}
\end{equation}
and because of the anomaly in (\ref{ghostvir}) it is not
nilpotent:
\begin{equation}
{d_{\rm gh}}\!\!^{2} = - \sum_{m>0} ( 2m + \small{\frac{26}{12}} m
(m^2 -1))c_{-m}\,c_{m}\;\;. \label{square}
\end{equation}

One can see that the ghost level operator ${\cal{L}}_{0} = \sum
m:c_{-m}\,b_m: $ is diagonalizable in ${\cal{C}}_{\rm gh}$ and it
commutes (in contrast to remaining operators of (\ref{vir_ghost}))
with the differential (\ref{ghostdiff}). Therefore the space
${\cal{C}}_{\rm gh}$ splits into a direct sum of finite
dimensional subspaces of fixed ghost level $N$ which are $d_{\rm
gh}$ invariant:
\begin{equation}
{\cal{C}}_{\rm gh} = \bigoplus_{N \geq 0} {\cal{C}}^{N}_{\rm
gh}\;\;. \label{ghostleveldec}
\end{equation}
Every subspace ${\cal{C}}^{N}_{\rm gh}$ of fixed level is graded:
\begin{equation}
\label{grading}
    {\cal{C}}^{N}_{\rm gh}= \bigoplus_{r} \,{\cal{C}}^{N,\,r}_{\rm
    gh}\;\;,\;\;\;d_{\rm gh} :{\cal{C}}^{N,\,r}_{\rm gh}
    \rightarrow
    {\cal{C}}^{N,\,r+1}_{\rm gh}\;\;,
\end{equation}
by half integral eigenvalues of the ghost number operator:
\begin{equation}
 {\rm gh} = \sum_{m\in {\mathbb{Z}}} :c_{-m}b_{m}\!: =  \sum_{m>0} (b_{-m}c_m - c_{-m}b_m)+
 \frac{1}{2}(c_0b_0 - b_0c_0)\;\;.
 \label{ghostnumber}
\end{equation}
The direct sum in the ghost number decomposition (\ref{grading})
ranges over finite and symmetric with respect to zero interval of
half-integers. The spaces ${\cal{C}}^{N,\,r}$ and
${\cal{C}}^{N,\,- r}$ are paired in a non-degenerate way with
respect to the scalar product defined by the vacuum normalization
condition and  formal conjugation rules (\ref{con}).

\subsection{The string differential spaces}

 The string differential space  is constructed as an appropriate
tensor product of the covariant string space $\cal{H}$
(\ref{covariant space}) and the ghost sector ${\cal{C}}_{\rm gh}$
described in the previous subsection. More precisely:
     \begin{equation}
    {\cal{C}} = \bigoplus_{N\geq 0}  \int d^{d}p \;{\cal{C}}^N (p)
    \;\;\;\;\;\;{\rm where}\;\;\;\;\;
    {\cal{C}}^N (p) = \bigoplus_{N'+ N'' = N } {\cal H}^{N'}(p)\otimes {\cal{C}}^{N''}_{\rm gh}\;\;.
 \label{totalcomplex}
 \end{equation}
 The decompositions of  the spaces ${\cal H}(p)$ of
(\ref{covariant space}):
\begin{equation}
{\cal H}(p) = \bigoplus_{N' \geq 0} {\cal H}^{N'}(p)\;\;.
\label{level matter}
\end{equation}
 into finite dimensional
eigensubspaces of string level operator $R_{\rm str} = L_{0} -
{\small \frac{1}{2\alpha}} p^2 - 2\beta $ were used in the
definition (\ref{totalcomplex}).
\\
 The string differential space is equipped with the grading inherited  from the ghost sector
 (\ref{grading}). The subspaces  of ${\cal{C}}^N (p)$  of opposite ghost number are
    mutually dual with respect to the canonical pairing induced on the tensor product (\ref{totalcomplex}).

    The  differential in this space is defined in the standard way \cite{GHV}:
\begin{equation}
D = \sum_{m\in {\mathbb{Z}}}\,(L_m - \delta_{m 0}a) \otimes c_{-m}
+ 1 \otimes {d_{\rm gh}}\;\;.\label{brst}
\end{equation}
Introducing the following operator:
\begin{equation}
\label{jot_plus}
 J^{(+)}_\xi = \sum_{m>0}r_\xi(m)c_{-m}c_{m}\;\;,
\end{equation}
where:
\begin{equation}
\label{parametry}
 r_\xi(m) = m(m^2 - \xi)\;,\;\; \xi = 1+\frac{24(1-a)}{(d-25)+
 48\beta}\;\;,
\end{equation}
one may check that the square of  the differential  equals to:
\begin{equation}
    \label{de_dwa}
 D^2 = \frac{1}{12}( (d-25)+ 48\beta )\,
J^{(+)}_\xi\;\;,
\end{equation}
i.e. the operator (\ref{brst}) is not nilpotent in general. \\
Lack
of the nilpotency (\ref{de_dwa}) implies that the differential is
also not invariant with respect to the ghost extended constraint
operators:
\begin{equation}
    L_m^{\rm tot} = \{\,b_m , D\,\} = L_m + {\cal L}_m - \delta_{m 0}a \;\;,\;\;m \in \mathbb{Z}\;\;.
\label{Ltot}
\end{equation}
From the general identity $[\,L_m^{\rm tot}, D\,] = [\,b_m , D^2\,]$
it immediately follows that:
\begin{equation}
    [\,L_m^{\rm tot}, D\,] = \frac{1}{12}( (d-25)+ 48\beta )\, r_\xi(m) \,c_m
    \;\;.
\label{inv}
\end{equation}

    Therefore the differential space
    $(\,{\cal{C}},D\,)$
is not the complex except of  the case when  the parameters
$\beta$ and $a$ take their critical values
(\ref{critical_values}). In this very special case the
differential space constructed above is equipped with nilpotent
differential operator and is used to be called the (standard)
BRST-complex.\\
    This standard structure was
introduced in \cite{hwang} for  $d=26$ Nambu-Goto critical string
model and in \cite{brstn} for critical massive string models with
Liouville modes. In  this last paper the cohomologies of critical
BRST-complexes were calculated and described in details.\\

\noindent
    In the remaining part of this chapter the complexes
 corresponding to anomalus square in (\ref{de_dwa}) and to
 generic cases of unitary non-critical massive string models (\ref{cont_series}) and
(\ref{disc_series}) will
 be introduced and analyzed.

\subsection{Anomalus complexes}
The constructions of this subsection can be applied to any
physical systems with anomalies.  For reasons explained in the
Introduction all further considerations will be concentrated on
string models however.

The space of the anomalus complex is defined in the simplest
possible way, namely  as the the maximal subspace of ${\cal C}$
(\ref{totalcomplex}) on which the canonical differential
(\ref{brst}) is nilpotent. It is clear that it coincides with the
kernel of the anomaly operator (\ref{de_dwa}).\\ The space of the
polarized anomalus complex is defined as the maximal subspace of
${\cal C}$ on which the differential (\ref{brst}) is invariant
(see (\ref{inv})) with respect to Gupta-Bleuler subalgebra
(\ref{phys}) of that formed by all ghost extended constraint
operators (\ref{Ltot}). \\ The remarks above are formalized in the
following
    \begin{definition}[Anomalus complexes]\
\begin{enumerate}
\item
    A differential space
    \begin{equation}
    \label{definition}
(\,{\cal{A}}\,,\,D \,)\;\;{\rm where}\;\;{\cal{A}} := \{\,\Psi \in
{\cal C}\;;\; D^2 \Psi = 0\,\} = {\rm ker}
 \,D^2
    \end{equation}
 is to be said anomalus BRST complex.
 \item
 A differential space
 \begin{equation}
    \label{definition_pol}
(\,{\cal{P}}\,,\,D \,)\;\;{\rm where}\;\;{\cal{P}} := \{\,\Psi \in
{\cal C}\;;\; [\,L_m^{\rm tot}, D\,] \Psi = 0\,;\; m \geq 0\, \}
    \end{equation}
 is to be said polarized anomalus BRST complex.
 \end{enumerate}
    \end{definition}
The definition (\ref{definition}) of the anomalus complex is by
all means consistent from the formal point of view: the kernel of
$D^2$ is preserved by the action of $D$.\\
    The space (\ref{definition_pol}) of polarized complex needs more care.  From
    the formulae (\ref{inv}) one immediately obtains
    $D^2 = \sum\, c_{-m}\,[\,L_{m}^{\rm tot},D\,]\,$.
    Consequently the space ${\cal{P}}$ is contained in ${\cal
    A}\,$ and $D^2\,{\cal{P}} = 0\,$. Hence the necessary condition for (\ref{definition_pol}) to
    be a complex is satisfied.\\
    It is however not obvious that the differential $D$ acts inside ${\cal P}$.
    It depends on the properties of the anomaly which are encoded
    in the coefficients $r_\xi(m)$ of the operator $ J^{(+)}_\xi$
    of (\ref{jot_plus}).\\
    In order to decide if $D {\cal P} \subset {\cal P}$ one
    should first notice that, in any case, the space ${\cal P}\,$ is preserved
    by the action of the ghost annihilation  operators:
    \begin{equation}
    \label{stable}
    {\rm if}\;\;\;\; \Psi \in {\cal P}\;\;\;\;{\rm then}\;\;\;\; b_m \Psi \in {\cal
    P}\;,\;\;m \geq 0\;\;.
    \end{equation}
 The above property follows from the general identity $\{\, b_k ,[\,L_m^{\rm tot}, D\,]\} =
 (m-k)L_{m+k}^{\rm tot} + [\,L_k^{\rm tot},L_m^{\rm tot}\,]\,$ and the fact
 that the subalgebra formed by $L_m^{\rm tot}\;,\;\;m \geq 0\,$ is anomaly free.\\
    Next, admitting  (for the moment) arbitrary real values of the parameter $\xi$ in the
    definitions (\ref{jot_plus}) and (\ref{parametry}) one may
    distinguish two (essentially) different cases.
   The first one corresponds to the non-degenerate anomaly and is defined by the following conditions on
   the parameter $\xi\,$: $r_{\xi}(m) \neq 0\,$ for all $m > 0$. In the second, opposite case, $\xi = s^2$
   for some integer $s > 0\,$ and the anomaly becomes
   degenerate at the $s$-th ghost mode: $r_{\xi}(s) = 0\,$.

 One is now in a position to prove the following
 \begin{lemma}\ \\
 The space $(\,{\cal{P}}\,,\,D \,)$ is a complex iff either the
 anomaly is non-degenerate or it is degenerate at the $1$-st ghost
 mode.
 \end{lemma}
{\it Proof} : According to what was said above  it is enough to
prove that the space ${\cal P}$ is preserved by the differential
$D$.\\
    In the non-degenerate case the conditions of the definition
    (\ref{definition_pol}) amount to $c_m \Psi = 0$ for all $m >
    0\,$. Hence ${\cal P}$ is generated by all ghost modes.
    The differentials of the $c_m$ ghost
    modes can be explicitly calculated ($m>0$):
    \begin{equation}
    \label{deceem}
\{\,D,c_m\,\} = - \sum_{k>0} (m + 2k) c_{-k} c_{m+k} +
\frac{1}{2}\sum_{k=1}^{m-1} (m-2k) c_{m-k}c_k - m c_0 c_m\;\;.
    \end{equation}
    The  formulae above  clearly implies that $\{\,D,c_m\,\}\mid_{\cal P} = 0$
    and $D{\cal P} \subset {\cal P}$ in non-degenerate case.\\
Assume now that the anomaly is degenerate at $s$-th mode:
    $r_{\xi}(s)= 0 \,$.
    Then the elements of ${\cal P}$ are defined by the
conditions $c_m \Psi = 0$ for all $m > 0$ except of $m = s$. In
this case the space ${\cal P}$ is generated by all ghost modes and
one anti-ghost  $b_{-s}$. Hence the state $\Psi = b_{-s}c_0 \Omega
(p)$, where $\Omega (p)$ is the ghost-matter vacuum located at
some momentum $p\,$, belongs to ${\cal P}\,$. Assuming that $D
\Psi \in {\cal P}$ one obtains
    $L_{-s}^{\rm tot}c_0 \Omega (p) \in {\cal P}$ as $D c_0\Omega (p) =
    0\,$. Due to (\ref{stable}) one gets
    $b_{1}L_{-s}^{\rm tot}c_0 \Omega(p) = (s +1) b_{-s + 1} c_0
    \Omega(p) \in {\cal P}\,$, which for $s>1$ yields
    contradiction. One is left with one possibility only, namely: $s=1$.
From the formulae (\ref{deceem}) one may easily see  that for
    $m>1$
    the differential $\{\,D,c_m\,\}$ does never contain $c_{1}$
ghost mode, which is not multiplied by the higher one. Therefore
$\{\,D,c_m\,\}\mid_{\cal P} = 0$ in this unique degenerate case.
$\;\;\Box$\\
\\
    It is worth to stress that for both series (\ref{cont_series}) and
    (\ref{disc_series}) of unitary massive string models one has
    $\xi \leq 1$. Therefore the polarized complexes always exist
    and moreover the only admissible degenerate case occurs in the
    continuous series for $a=1$.\\
    It has to  be also mentioned that the spaces introduced in the
    definitions (\ref{definition}) and (\ref{definition_pol})
    depend on the parameter $\xi$ (\ref{parametry}) of the string models. For the sake of simplicity
    this dependence will never be marked in the notation
    explicitly.\\
\\
 In order to say more on the structure of the anomalus complex it should be noted that
 it  can be decomposed in the same way as
    total string differential space (\ref{totalcomplex}):
   \begin{equation}
   \label{anomlevel}
{\cal{A}}
    = \bigoplus_{N\geq 0} \int d^{d}p \;{\cal{A}}^N (p)\;,
   \end{equation}
 i.e. into a direct sum/integral of subcomplexes supported by the subspaces
  of fixed momentum and level. The polarized complex admits an
  analogous decomposition.\\
\\
  Due to the fact that the anomaly (\ref{de_dwa}) acts in the matter sector as
  multiplication by (non-zero) number, the fixed level and momentum
  components in (\ref{anomlevel}) can be factorized in the same way as in full differential space (\ref{totalcomplex}):
\begin{equation}
\label{factorization}
 {\cal{A}}^N (p) =
\bigoplus_{N' + N'' = N } {\cal H}^{N'}(p)\otimes
{\cal{A}}^{N''}_{\rm gh}\;\;,
\end{equation}
but with restricted ghost factors this time:
    \begin{equation}
    \label{ghostfactor}
{\cal{A}}^{N''}_{\rm gh}\; =\; \{\, c \in {\cal C}_{\rm gh}^{N''}
\,;\;\;J^{(+)}_\xi c = 0 \,\}\; =\; {\rm ker}\,J^{(+)}_\xi\;\;.
    \end{equation}
 Since the ghost number operator (\ref{ghostnumber}) preserves the kernel of
 $D^2 \!\!:\;[\,{\rm gh}\,,\,D^2] = 2 D^2$ the spaces ${\cal{A}}^N
 (p)$ admit the grading induced from (\ref{grading}) i.e. every
 element in the kernel of $J^{(+)}_\xi$ can be decomposed into
 components of fixed degree. Hence:
    \begin{equation}
    \label{anomalusgrading}
    {\cal{A}}^N (p) = \bigoplus_{r} \,{\cal{A}}^{N,\,r}(p)\;\;\;\;
    {\rm and}\;\;\;\;D :{\cal{A}}^{N,\,r}(p)
    \rightarrow
    {\cal{A}}^{N,\,r+1}(p)\;\;.
    \end{equation}

 Not all the properties of the original grading (\ref{grading})
are inherited from the total ghost sector.  It will be made
evident that  all  ghost numbers lower than $-\frac{1}{2}$ for
non-degenerate anomaly and below $-\frac{3}{2}$ in degenerate case
are excluded by (\ref{ghostfactor}).  For this reason the
canonical pairing induced from the total complex becomes
degenerate on ${\cal{A}}$ and $\cal{P}$.
\\
\\
Depending on the case one introduces the complementary operator:
\begin{equation}
\label{jot_minus}
  J^{(-)}_\xi= \left\{
\begin{array}{ll}
\; {\sum\limits_{m>0}}\frac{1}{r_\xi(m)}\,b_{-m}b_{m} & ;\;\;\xi <
1 \cr \;\; \cr \;\sum\limits_{m>1}\frac{1}{r_\xi(m)}\,b_{-m}b_{m}
& ;\;\;\xi = 1 \;\;.
\end{array}
\right.
\end{equation}
    The operator $ J^{(-)}_\xi$ neither commutes with $D$ nor it does with
     $D^2$. Therefore $ J^{(-)}_\xi$ does not preserve the space ${\cal A}$
    underlying the anomalus complex (\ref{definition}).
    Nevertheless it is very useful in closer identification of
    the structure of the  subspaces ${\cal{A}}^N(p)$ of the  anomalus
    complex.\\
The commutator  of $J^{(-)}_\xi$ with the anomaly operator
$J^{(+)}_\xi$ is almost $\xi$ independent, more precisely:
\begin{equation}
\label{sldwa}
 [\,J^{(+)}_\xi ,\,J^{(-)}_\xi \,] = J^{(0)} := {\rm gh} - {\rm gh}_{(0)} -  \delta_{(\xi-1)}{\rm
 gh}_{(1)}\;\;,
\end{equation}
    where ${\rm gh}$ is the ghost number operator of
 (\ref{ghostnumber}) and ${\rm gh}_{(m)}$ denotes its $m$-th
 mode component. \\
From the commutation relations:
\begin{equation}
\label{com}
 [\,J^{(0)},\,J^{(\pm)}_\xi\,] = \pm 2 J^{(\pm)}_\xi\;\;,
\end{equation}
 it follows that $J^{(\pm)}_\xi$  together with $J^{(0)}$ form the
structure of $sl(2,\mathbb{C})$ Lie algebra. \\
 Hence, the subspaces ${\cal{A}}^N(p)$ of the anomalus
complex  consist of all elements of ${\cal{C}}^N(p)$, which are of
highest weights (coprimitive cochains \cite{Wells}) with respect
to the above $sl(2,\mathbb{C})$-Lie algebra.

    It is well known
(e.g. \cite{bourbaki}) that the commutation relations
(\ref{sldwa}) imply that the highest weight vectors
 have  non-negative weights  with respect to
${ {J}}^{(0)}$. The operator ${ {J}}^{(0)}$ does not feel the
ghost zero-modes in non-degenerate case and in addition
$c_{-1}$, $b_{-1}$ ghost modes for $\xi = 1$. \\
From the relation (\ref{com}) it follows that the decomposition of
the anomalus complexes at fixed momentum and level with respect to
the ghost number can be summarized in the following form:
\begin{equation}
\label{anomalus_grading}
  {\cal{A}}^N(p)= \left\{
\begin{array}{ll}
\;\bigoplus\limits_{r\geq -\frac{1}{2}}\, {\cal{A}}^{N, r}(p)  &
;\;\;\xi < 1\;\; \cr \;\; \cr \;\bigoplus\limits_{r\geq
-\frac{3}{2}}\, {\cal{A}}^{N, r}(p) & ;\;\;\xi = 1\;\;\;.
\end{array}
\right.
\end{equation}
\\
\\
It is worth to make a comment on the differences of the polarized
and anomalus co-chains. From the conditions (\ref{definition_pol})
and the identity (\ref{inv}) it follows that the polarized complex
is generated by ghost modes $c_{-m}\;,m \geq 0$ and in addition
$b_{-1}$ anti-ghost in the  degenerate case. The anomalus complex
contains much more co-chains. Besides of the above, polarized
elements, there are co-chains containing ghost/anti-ghost
clusters. The clusters are generated by:
\begin{eqnarray}
\label{clusters}
   G_{-r,-s} &:=& [\,J^{(-)}_\xi,( c_{-r}+ c_{-s})\,]\,(c_{-r} +
    c_{-s}) - G_{-2r} - G_{-2s} \;\;\;\;\;\;\;{\rm
    and}\;\;\;\cr
     G_{-2m}\;\, &:=& [\,J^{(-)}_\xi, c_{-m}\,]\,c_{-m}
    \;,\;\;\;\; m, r \neq s > 0\;\;.
\end{eqnarray}
From the definition (\ref{clusters}) it immediately follows that
these elements (of ghost number zero) are scalars with respect to
$sl(2,\mathbb{C})$-Lie algebra generated by the anomaly i.e.
$[\,J^{(\pm)}_\xi, G_{(\cdot)}\,] = 0\,$.\\
\\
As in the critical case there are several complexes associated
with the anomalus BRST complex (\ref{definition}) and its
polarized subcomplex (\ref{definition_pol}). Although only the
cohomologies of the polarized subcomplex will be calculated and
identified, it is worth to define the associated complexes for the
full anomalus complex. The corresponding polarized structures can
be obtained by simple restriction to the subspace defined in
(\ref{definition_pol}).\\
    First of all one  introduces the on-mass-shell complex:
 \begin{equation}
{\cal{A}}_0 := {\rm ker }\,( L_0^{\rm tot}|_{{\cal{A}}}) =
\bigoplus_{N\geq 0}\int_{S_{N}}d\mu^N (p)\,{\cal{A}}^N (p)\;,
\label{massshellcomplex}
\end{equation}
which contains the roots of all non-trivial cohomology classes of
the anomalus complex (\ref{definition}). The elements of
${\cal{A}}^N (p)$ are supported on the mass-shells $S_N$ defined
by the equations:
\begin{equation}
 p^2 = - 2\alpha\left( N^{\rm str} +
 N^{\rm gh} + 2\beta - a \right)\;\;,\;\;\; N = N^{\rm str} +
 N^{\rm gh}\;\;.
 \label{massshell}
\end{equation}
The property that any non-trivial cohomology class of
(\ref{definition}) stems from the element of
(\ref{massshellcomplex}) follows  from the fact that the operator
    $L_0^{\rm tot} = {\cal{L}}_{0} + L_{0} - a$
is diagonal in  the space (\ref{definition}) and moreover it is
cohomologically trivial:
    $L_0^{\rm tot} = \left\{ b_0 , D\right\}$.
Consequently any $D$-closed element outside its kernel is exact
(e.g. \cite{brstn}). Hence the cohomologies of  the on-mass-shell
complex $(\,{\cal{A}}_0\,,\, D\,)$ are identical with these of
(\ref{definition}).

Once the complex is restricted to  on-mass-shell cochains and it
is observed that the ghost $c_0$ (corresponding to the kinetic
operator $L_0^{\rm tot}$) does not
contribute to the mass spectrum it is natural to get rid of it and
to introduce a relative complex:
\begin{equation}
{\cal{A}}_{\rm rel}(p) := \left\{ \,\Psi\in {\cal{A}}_0(p)
\;\;\;;\;\; b_0 \Phi = 0 \,\right\}\;\;, \;\;{\cal{A}}_0 (p) =
{\cal{A}}_{\rm rel}(p) \oplus c_0\, {\cal{A}}_{\rm rel}(p)\;\;,
\label{relativecomplex}
\end{equation}
with the differential:
\begin{equation}
 D_{\rm rel} := D - L_0^{\rm tot} c_0 - M b_0 \;\;;\;\;M =
-2\sum_{m>0} m c_{-m}c_m = \{ D\,,\,c_0\}\;\;.
\label{relativediff}
\end{equation}
The square of relative differential can be calculated on the total
differential space (\ref{totalcomplex}) and there it gives:
\begin{equation}
\label{relative_square}
 D_{\rm rel}^2 =  \frac{1}{12}\left( (d-25)+ 48\beta \right)1
J^{(+)}_\xi- M L_0^{\rm tot}\;\;.
\end{equation}
The first term on the right hand side vanishes on anomalus
cochains of ${\cal{A}}$ while the second one is zero on the
mass-shell. Hence $D_{\rm rel}$ is nilpotent on ${\cal{A}}_{\rm
rel}$.

     Since the ghost zero mode $c_0$ is absent in ${\cal{A}}_{\rm
rel}$ it is natural and convenient  to introduce the integral
grading in ${\cal{A}}_{\rm rel}$ by  eigenvalues of relative ghost
number operator:
\begin{equation}
\label{relative_ghost}
    {\rm gh}_0 =  \sum_{m>0} (b_{-m}c_m -
c_{-m}b_m)\;\;.
\end{equation}
This change of grading amounts to the shift $r \rightarrow
r+\frac{1}{2}$ in the degrees (\ref{anomalusgrading}) of all
cochains from ${\cal{A}}_{\rm rel}$.\\
\\
 The anomalus relative complex $ {\cal{A}}_{\rm rel}(p)$ admits (as in critical case \cite{brstn})
richer grading structure than the one defined by the eigenvalues
of the relative ghost number operator(\ref{relative_ghost}):
\begin{equation}
\label{bigrading}
  {\cal{A}}_{\rm
 rel}^r(p) = \bigoplus_{a-b = r}{\cal{A}}^a_b\;\;,
\end{equation}
where $a,b$ denote the ghost $c_{-m}$ excitation degree and
anti-ghost $b_{-m}$ excitation degree respectively.\\ The
possibility to decompose any highest weight vector of
${\cal{A}}_{\rm
 rel}^r(p)$ into
bigraded components which are also of highest weights follows from
the fact that the following  operator:
\begin{equation}
\label{central}
 \overline{{\rm gh}}_0 = \sum_{m>0} (b_{-m}c_m + c_{-m}b_m)\;\;,
\end{equation}
which counts the sums $a+b$, is central with respect to the
$sl(2,\mathbb{C})$ algebra of (\ref{sldwa}).

The bigraded structure (\ref{bigrading}) makes it sensible and
useful to decompose the relative differential accordingly $ D_{\rm
rel} = {\cal D}\;\; +\;\; \overline{{\cal{ D}}}\;$:
\begin{eqnarray}
{\cal D} &=& \sum_{m>0} L_{-m}\otimes c_m \,+\, \sum_{m>0} c_m
\tau_{-m} \,+\,
\partial\;\;,\cr
 &~~& \cr
\overline{{\cal D}} &=& \sum_{m>0} L_{m}\otimes c_{-m}\,+\,
\sum_{m>0} c_{-m} \tau_{m}\,+\, \overline{\partial}\;\;,
\label{differentials}
\end{eqnarray}
where:
    $$
\partial = -\frac{1}{2}\sum_{m,k >0 }(m-k)b_{-k-m}c_kc_m \;, \;\;\;\;\;
 \overline{\partial} = -\frac{1}{2}\sum_{m , k>0
}(m-k)c_{-m}c_{-k}b_{k+m}\;\;,
    $$
and:
    $$
 \tau_{m} = \sum_{k>m}(m+k)b_{m-k}c_{k} \;,\;\;\;\;\;
 \tau_{-m} = \sum_{k>m}(m+k)c_{-k}b_{k-m}\;\;\;;\;
 m>0\;\;.
    $$
It is remarkable\footnote{and in fact it seems to be universal
within the wide class of constrained systems} property that
independently of the presence of the anomaly the differentials
${\cal D}$ and $\overline{{\cal D}}$ are nilpotent on the whole
ghost extended string space (\ref{totalcomplex}). The anomaly is
hidden in their anticommutator:
    \begin{eqnarray}
    \nonumber
{\cal D}^2 = 0\;\;\;&,&\;\;\;\overline{{\cal D}}^2= 0 \;\;,\cr
 &~~& \cr
 {\cal D}\,\overline{{\cal D}}+\, \overline{\cal D}{{\cal D}} &=&
\frac{1}{12}\left( (d-25)+ 48\beta \right)1 \otimes J^{(+)}_\xi- M
L_0^{\rm tot}\;\;,
    \end{eqnarray}
and both differentials obviously commute with $J^{(+)}_\xi$. \\
    The above relations  define the structure of  $sl(1 | 1;\mathbb{R})$
Lie superalgebra. This structure was already used within slightly
different, although related approach to quantum constrained
systems in \cite{GB}.\\
\\
 The bigraded structure\footnote{This structure is in fact induced from the one
 of the total differential space \cite{brstn}.}
  of  the anomalus (relative) complex
can be most clearly summarized in the form of the following
diagram:
\begin{equation}
\label{diagram}
\begin{CD}
    @.      \vdots  @.  \vdots  @.  \\
    @.      @VV{\cal D}V            @VV{\cal D}V        @.      \\
\cdots @>{\overline{{\cal D}}}>>    {\cal{A}}^a_{b} @>{\overline{{\cal D}}}>>
{\cal{A}}^{a+1}_{b} @>{\overline{{\cal D}}}>>   \cdots\\
@.      @VV{\cal D}V            @VV{\cal D}V        @.      \\
\cdots @>{\overline{{\cal D}}}>>    {\cal{A}}^a_{b-1}   @>{\overline{{\cal
D}}}>>   {\cal{A}}^{a+1}_{b-1}   @>{\overline{{\cal D}}}>>   \cdots\\
@.      @VV{\cal D}V            @VV{\cal D}V        @.      \\
    @.      \vdots  @.  \vdots  @.
\end{CD}
\end{equation}
The form (\ref{differentials}) of ${\overline{{\cal D}}}$ and $
{{\cal D}}$ differentials suggests that horizontal maps impose the
constraints on the states, while the vertical maps implement the
gauge transformations.\\
    The comment on the corresponding structure of the polarized
    complex must be added here. In the case of non-degenerate
    anomaly the only admissible anti-ghost degree of the polarized
    elements is zero i.e. ${\cal P}_b^a = 0\,,\;\;b>0\,$. In the
    degenerate case the co-chains of anti-ghost degree equal to
    one (created by $b_{-1}$ mode) are present and there is
    residual gauge symmetry generated by $L_{-1}^{\rm tot}$
    operator.\\
    This structure is most clearly visible in the form of the
    relative differential restricted to the polarized complex.
    Since $\partial\mid_{\cal P}\equiv 0$, one has on ${\cal P}$:
\begin{equation}
\label{polarized_relative}
    D_{\rm rel} = {\overline{{\cal D}}} + \delta_{(\xi-1)}(L_{-1} +
    \tau_{-1})c_1\;\;,
\end{equation}
where $\tau_{-1}= \sum_{k>0}(k+2)c_{-k-1}b_k $ is that of
(\ref{differentials}).
\\
\\
The bigraded structure of the polarized relative complex, as one
will see in the next section, appears to be very useful in the
computation of polarized cohomologies and in identification of
their representatives in terms of physical states mentioned in the
first chapter.

\section{Cohomologies of the polarized BRST complex}

This chapter is devoted to the computation of the cohomology
spaces of polarized anomalus complexes introduced in the previous
section. The identification of the representatives of non-zero
cohomology classes with the elements of the physical space
(\ref{phys}) of ~"old covariant" approach will be given explicitly
too.\\
\\
 The cohomology
spaces of the total complexes (\ref{definition_pol})  may be
formally reconstructed as a direct sums/integrals out of
cohomology spaces:
\begin{equation}
H^r(p) :=\frac{Z^r(p) }{B^r(p)}\;\;;\;\;Z^r(p) := {\rm ker }\,
D|_{\,{\cal P}^r (p)}\;,\;\; B^r(p) = {\rm im}\, D|_{\,{\cal P}^{
r-1}(p)} \;\;, \label{totalcohomologyspaces}
\end{equation}
of the corresponding  subcomplexes located at fixed momentum and
level.
\\
   The convention to denote the spaces of cocycles by the
root letter $Z$ and those of coboundaries by the root letter  $B$
will be kept throughout this paper. \\
 The  relative cohomology spaces located at fixed momentum are defined
    in exactly analogous way.
    It is only worth to introduce the bigraded cohomology spaces
corresponding to the structure illustrated in the diagram
(\ref{diagram}):
\begin{equation}
\overline{\cal H}^a_b(p) =\frac{ {\overline{\cal
Z}}^a_b(p)}{{\overline{\cal B}}^a_b(p)}\;\;\;, \;\;\;{\cal
H}_b^a(p) =\frac{{\cal Z}_b^a (p) }{{\cal B}^a_b(p) }\;\;.
\label{bigradedcohomology}
\end{equation}
The spaces $\overline{\cal H}^a_b(p)$ and ${\cal H}^a_b(p)$ denote
the cohomology spaces of ${\overline{{\cal D}}}$ and $ {{\cal D}}$
respectively.\\ It should be mentioned that for non-degenerate
anomaly the co-boundaries of ${\cal D}$ are zero and all the
elements of ${\cal P}_0^a$ are ${\cal D}$ co-cycles. Hence ${\cal
H}_0^a(p) = {\cal P}_0^a(p)\,$ in this case.

The order of calculating the cohomologies of respective anomalus
complexes will exactly opposite to the one they were introduced in
the previous section.\\
 First of all   it will be shown that an analog of Dolbeaut-Grotendick
 vanishing theorem for bigraded cohomologies is true. Using this
 result it will be proved that also the relative classes do vanish for non zero
 (positive in anomalus case) ghost numbers.
 The transparent representation of the non trivial
 relative cohomology spaces will be given in terms of bigraded cocycles.
 This will in turn allow one to
 identify the relative classes with the physical states of ~"old covariant
 formalism" mentioned in the first chapter. Finally the absolute cohomologies
will be reconstructed out of relative ones.\\
\\
    It is worth to mention that nothing like Poincare-Serre
    duality \cite{brstn} can be used in the anomalus case. This is a consequence of the fact that there
    are no anomalus cochains of arbitrarily low negative ghost number
    and the natural pairing induced from
    full differential space ${\cal C}$ of (\ref{totalcomplex}) is
    highly degenerate on the anomalus complex as well as on its polarized subcomplex.
    Therefore one cannot use the canonical duality properties of the
    respective cohomology spaces.
    For this reason the cohomologies of ${\overline{{\cal D}}}$ and $ {{\cal
    D}}$ should be calculated independently. Due to the fact that
    the differential ${\cal D}$ has residual form of
    (\ref{polarized_relative}) it is not necessary to know its
    cohomologies in order to prove the Vanishing Theorem for the
    relative classes of non-zero ghost number. It is also not
    needed for the identification of the representatives of
    non-vanishing classes with the physical states (\ref{phys})
     of ~"old covariant approach".

As in \cite{brstn}, \cite{BMP} and \cite{BT} the method of
 filtered complexes can be successfully  used also here. For the sake of
 completeness the steps of reasoning will be repeated.

First, one has to choose the momentum supporting the space of
cochains. Once a non-zero momentum $p$ is fixed one may always
find an adapted light cone basis\footnote{Consisting of two light
like vectors $ e_{\pm}^2 = 0 \,,\, e_+\cdot e_- = -1$ and an
orthonormal basis $\{ e_i \}_{i = 1}^{d-2} $ of Euclidean
transverse space. } $\{\, e_\pm,e_i \,\}_{i=1}^{d-2}$ of the
momentum space such that
 $p^+ := e_+ \cdot p \neq 0$. The
Virasoro operators of the string sector (\ref{virasoro}) written
in the light-cone basis adapted to the momentum $p$ take the form:
\begin{eqnarray}
L_m = &-& {\small{\frac{1}{\sqrt{\alpha}}}} p^+ a_m^- + \cr &+&
L_m^{\rm tr} + L_m^{\rm Li}\;\; - \sum_{n\neq0 \,\,m+n\neq 0}\;
a^+_{n+m}a^-_{-n} - {\small{\frac{1}{\sqrt{\alpha}}}}p^-
a_m^+\;\;. \label{filtervir}
\end{eqnarray}
The symbols $L_m^{\rm tr}$ and $ L_m^{\rm Li} $ denote the
operators given by  standard expressions (\ref{virasoro}) in
transverse and Liouville modes respectively. \\
     Next, one can
introduce  an additional gradation in the spaces
(\ref{definition_pol}) by assigning the following filtration
degrees to the elementary matter/ghost modes:
\begin{eqnarray}
{\rm deg} (a^-_m ) &=& - 1 \;\;,\;\;{\rm deg }(b_m) = -1 \cr
 {\rm
deg }( a^+_m ) &=& + 1 \;\;,\;\; {\rm deg} ( c_m ) =  + 1\cr
 {\rm deg
}(a^i_m) &=& \,\;\;0\;\;,\;\;{\rm deg }(u_m) = \;\;0 \;;\;\; m \in
{\mathbb{Z}}\setminus \{0\} \;\;,\; 1\leq i\leq d-2\;\;.
\label{filtration}
\end{eqnarray}
 The spaces ${\cal{P}}^a_b(p) $ of bigraded complex
(\ref{bigrading}) decompose into direct sums of filtration
homogeneous components: ${\cal{P}}^a_b(p) = \bigoplus_{f}
{\cal{P}}^a_{b ; f}(p)$. According to this decomposition  the
differential $\overline{{\cal D}} $  decays into three parts:
\begin{equation}
    \overline{{\cal D}} = \overline{{\cal D}}_{(0)}
    + \overline{{\cal D}}_{(1)} + \overline{{\cal D}}_{(2)} \;,\;\;
\overline{{\cal D}}_{(0)}
    = - \sum_{m > 0}\,{\small{\frac{1}{\sqrt{\alpha}}}} p^+ a_m^-\;
    c_{-m}\;\;.
\label{filtersplit}
\end{equation}
The operators $ \overline{{\cal D}}_{(i)}\;,  i = 1,2$ can be
easily read off from (\ref{differentials}) and (\ref{filtervir})
but their explicit form is not needed. It only worth to mention
that they both act by rising the filtration degree:
\begin{equation}
\label{filteraction}
    \overline{{\cal D}}_{(i)} :{\cal{P}}_{b ;f}^a(p) \rightarrow {\cal{P}}_{b ;
f +i}^{a+1}(p)
   \;\;.
\end{equation}
Out of (\ref{filtersplit}) only the component $ \overline{{\cal
D}}_{(0)}$ of filtration degree zero will be in explicit use. It
is nilpotent and defines the cohomology space $ \overline{\cal
H}_{b;f}^a(p)$ localized at fixed filtration degree $f$.
\\
\\
Equipped with the above tools, by a slight modification of the
arguments used in \cite{brstn}, one may prove the following
important
    \begin{lemma}
$$ \overline{\cal H}_{b;f}^a(p) = 0\;\,;\;\;\;  a > 0\;\,,\;\;
p\neq 0 \;\;. $$
    \end{lemma}
{\it Proof} :
    The operator of filtration degree zero:
    $$\overline{{\cal R}}
=\sum_{m>0}( c_{-m} b_m - \frac{1}{m}a^+_{-m}a^-_{m})\;\;, $$
counts the total degree of $c_m$ ghost and $a_m^+$ string
excitations hence it is non-negative.  From the statement
(\ref{stable})
    it follows that it acts within the spaces ${\cal{P}}^a_{b}$ of
    polarized cochains contained in(\ref{bigrading}). The operator $\overline{{\cal
    R}}$ is exact:
    \begin{equation}
\label{homo}
    \overline{{\cal R}} = \left\{\, \overline{{\cal
    D}}_{(0)}\,,\,{\cal{\overline
    K}}\,\right\}\;,\;\;\;{\rm where}
    \;\;\;\overline{{\cal K}} = {\small
\frac{\sqrt{\alpha}}{p^+}}\sum_{m>0}\,\frac{1}{m}
    a^+_{-m}b_m\;\;,
    \end{equation}
on the whole representation space ${\cal{C}}_{b}^\bullet$ and again
due to (\ref{stable}) also on ${\cal{P}}_{b}^\bullet\,$.\\
     Let
$\Psi$ be $\overline{{\cal D}}_{(0)}$ closed cochain from
${\cal{P}}^\bullet_b(p)$ . It may be assumed that $\Psi$ is an
eigenstate of $\overline{{\cal R}}$: $\overline{{\cal R}}\Psi =
s\Psi$. From the relation (\ref{homo}) it follows that
    \begin{equation}
    \label{exact}
    \Psi = \frac{1}{s}\overline{{\cal D}}_{(0)}\overline{{\cal K}}\Psi\;\;,
    \end{equation}
    for any cocycle of  $\overline{{\cal R}}$-degree $s\neq 0$.
Consequently all $\overline{{\cal D}}_{(0)} $ closed states not in
the kernel of $ \overline{{\cal R}}$ (in particular those with $a
> 0$) are exact. $\;\;\Box$\\
\\
Making use of the above statement it is possible to prove the
strict counterpart of the Dolbeaut-Grotendieck lemma of classical
complex geometry \cite{Wells} on vanishing of bigraded
cohomologies (\ref{bigradedcohomology}) of $\overline{{\cal D}}$.

\begin{lemma}[Dolbeaut - Grotendieck]
\begin{eqnarray}
\overline{\cal H}_b^a(p) = 0 \;\,;\;\; a
> 0\;\,,\;\; p \neq 0\;\;. \nonumber
\end{eqnarray}
\end{lemma}
{\it Proof} :
 Since the filtration degree is bounded at any fixed level the
cochain of bidegree $(a,b)$ can be decomposed into finite sum of
homogeneous components with respect to the filtration degree:
$\Psi^a_b = \sum_{i\geq m}\,\Psi^a_{b ;\, i}$. The equation
${\overline{\cal D}} \Psi^a_b = 0$ written in terms of
(\ref{filtersplit}) implies a chain of equations for homogeneous
constituents and in particular $ \overline{{\cal D}}_{(0)}
\Psi^a_{b;\, m} = 0$ for the component of lowest filtration
degree. Since the cohomologies of $\overline{{\cal D}}_{(0)} $ are
trivial $ \Psi^a_{b;\, m} = \overline{{\cal D}}_{(0)}F^{a-1}_{b
;\, m}$ (Lemma 3.1). The lowest filtration component of an
equivalent element $\Psi'^a_b = \Psi^a_b - {\overline{\cal
D}}F^{a-1}_{b;\, m}$ is of degree at least $m+1$. The procedure
repeated appropriately many times leads to the conclusion that $
\Psi^a_{b}= {\overline{\cal D}}\Phi^{a-1}_{b} $ for some
$\Phi^{a-1}_{b}$.$\;\;\Box$\\
\\
The above lemma  implies directly the Vanishing Theorem for
relative cohomology and provides a convenient description of
non-vanishing classes at ghost number zero in terms of bigraded
cocycles.

\begin{theorem}[Vanishing Theorem]
$$
\begin{array}{l} 1)\;\;\; H^r_{\rm rel} (p) = 0 \;;\;\;r \ne
0\;,\;\;\;p\ne
0~~(on-shell)\;\;\;\;\;\;\;\;\;\;\;\;\;\;\;\;\;\;\;\;\;\;\;
\;\;\;\;\;\;\;\;\;\;\;\;\;\;\;\;\;\;\cr \cr 2)\;\;\;
 H^0_{\rm rel} (p) \sim   \left\{
\begin{array}{ll}
{\overline{\cal Z}}^0_0(p) & ;\;\;\;\;\xi < 1 \cr \;\; \cr
{\overline{\cal Z}}^0_0(p)\,/\,{\cal D}\,(b_{-1}{\overline{\cal
Z}}^0_0(p)) & ;\;\;\;\;\xi = 1\;\;.
\end{array}\;
\right.
\end{array}
$$
\end{theorem}
{\it Proof} : 1)  Take  $ \Psi^{r}\in Z^r_{\rm rel}(p)$. In the
case of  $r>0$ $ \Psi^{r}$ has the following bidegree
decomposition: $ \Psi^{r} = \Phi_0^r +
\delta_{(\xi-1)}\Phi_{1}^{r+1}$. The equation for $\Psi^r$ to be a
relative co-cycle reads:
    $$
    \delta_{(\xi-1)}\overline{{\cal D}}\Phi_{1}^{r+1} +
    \overline{{\cal D}}\Phi_0^r + \delta_{(\xi-1)}{{\cal
    D}}\Phi_1^{r+1} = 0\;\;.
    $$
In particular it should be  $\overline{{\cal D}}\Phi_{1}^{r+1} =
0\,$. Taking into account Dolbeaut - Grotendieck vanishing lemma
for $\overline{\cal H}_b^a(p)$ one has the solution $\Phi_1^{r+1}
= \overline{{\cal D}}F_{1}^{r}$ for some co-chain $F_{1}^{r}$ of
lower degree. Taking an equivalent element $ \Psi^r \sim
\tilde{\Psi}^{r} = \Psi^r - \delta_{(\xi -1)}D_{\rm rel} F_1^{r} =
\tilde{\Phi}_0^{r}$ one gets $\overline{{\cal
D}}\tilde{\Phi}_0^{r} = 0$. Hence again due to
Dolbeaut-Grotendieck vanishing lemma $ \tilde{\Phi}_0^r =
\overline{{\cal D}}\tilde{F}_{0}^{r-1} = D_{\rm
rel}\tilde{F}_{0}^{r-1}\,$. Finally $\Psi^r \sim D_{\rm
rel}\tilde{F}_{0}^{r-1}\,$ and $H^r_{\rm rel} (p) = 0$ for
$r>0$.\\
    For the negative ghost number (only $r=-1$ has to be
taken into account and only in the case of $\xi = 1$) one should
take $\Psi^{-1} = b_{-1}\Phi^0_{0}\,$. The condition for
$\Psi^{-1}$ to be a co-cycle with respect to $D_{\rm rel}$ amounts
to $\overline{{\cal D}}\Phi_0^0 = 0\,$ and $\{{\cal D},b_{-1}\}\,
\Phi_0^0 = 0 $. The first equation follows from the fact that
$\overline{{\cal D}}$ anti-commutes with the first anti-ghost mode
$b_{-1}$ (which is not in fact true for the higher ones) and means
that $\Phi_0^0$ is (up to ghost vacuum factor) the physical state
in the sense  of (\ref{phys}) i.e. $L_n \,\Phi_0^0 =0\,$ for $n
>0\,$. On the other hand from (\ref{polarized_relative}) it
follows that the second equation reduces to $L_{-1}\Phi_0^0 = 0$.
Hence also $L_1 L_{-1} \Phi_0^0 = [\,L_1,L_{-1}]\Phi_0^0 = 0\,$ and
as a consequence of the mass-shell condition (\ref{phys}) and
vanishing of the anomaly at the first ghost/antighost mode one
obtains $ L_0 \Phi_0^0 = \Phi_0^0 = 0\,$.
\\
    2) Taking $\Psi^0 = \Phi_0^0 + \delta_{(\xi-1)}b_{-1}\Phi_0^1$
    of the most general bi-degree decomposition in the polarized
    complex ${\cal P}$ and imposing $D_{\rm rel} \Psi^0 = 0$, one obtains the following conditions for
    its homogeneous components:
    $$
    \overline{{\cal D}}\Phi_0^1 = 0 \;\;\;\;{\rm and}\;\;\;\; \overline{{\cal
    D}}\Phi_0^0 + \delta_{(\xi-1)}{{\cal D}} (b_{-1}\Phi_0^1) = 0\;\;.
    $$
Again from Dolbeaut-Grotendieck vanishing lemma one obtains
$\Phi_0^1 = \overline{{\cal D}} F_0^0\,$ for some $F_0^0$ from
${\cal P}_0^0$. Taking an equivalent element $\Psi^0 \sim
\tilde{\Psi}^0 = \Psi^0 + \delta_{(\xi-1)}D_{\rm rel} (b_{-1}
F_0^0) = \tilde{\Phi}_0^0\,$ one obtains
    ${\cal D}\tilde{\Phi}_0^0 = 0$  .
    Hence any non trivial class from $ H^0_{\rm
rel} (p)$ stems from the co-cycle of ${\overline{\cal
Z}}^0_0(p)\,$. Further identification of ${\overline{\cal
Z}}^0_0(p) $ cocycles amounts to $ \Psi_0^0 - \tilde{\Psi}_0^0
=\delta_{(\xi-1)}{{\cal D}}(b_{-1} F^0_0)$ with $F_0^0$ being the
physical state i.e. ${\overline{\cal D}} F^0_0 = 0$. $\;\;\Box$\\
\\
The result above and the statement 2) in particular, allows one to
identify the representatives of non vanishing cohomology classes
with the physical states (\ref{phys}) of "old covariant
quantization". It is enough to notice (as it was already done in
the proof of the above Theorem) that the condition for $
\Psi^0_0(p) = \upsilon(p)\otimes\omega\,; \;\upsilon(p) \in {\cal
H}(p) $ to be a cocycle reads:
    $$
    0 = \overline{{\cal D}}\Psi^0_0(p)
= \sum_{m>0}\,L_{m}\upsilon(p)\otimes c_{-m}\omega\;\;,
    $$
and consequently:
\begin{equation}
{\overline{\cal Z}}^0_0(p)= {\cal H}_{\rm phys} (p) \otimes
\omega\;\;, \label{zetzerozero}
\end{equation}
where ${\cal H}_{\rm phys} (p)$ is the subspace (\ref{phys}) of the
original covariant space (\ref{covariant space}) of string states.
\\ In non-degenerate case $(\xi < 1)$ there is no any additional equivalence
(gauge) relation between them. For $\xi=1$ one obtains residual
gauge symmetry generated by $L_{-1}$ constraint operator.
    This appearance is tightly related to the fact
that the Gupta-Bleuler subalgebra, used to define the physical
states (\ref{phys}), supplemented by $L_{-1}$, is first of all
anomaly free (as the corresponding constraints are of first class
on the space of the total complex (\ref{totalcomplex})) and in
addition, it forms the parabolic subalgebra of (\ref{virasorocom})
in the sense of \cite{bourbaki}.

    The above situation should be compared and contrasted with the one
 encountered in the critical massive string model. As it
    was demonstrated in \cite{brstn} the relative cohomology space
    had the form of the quotient:
    $$
    H^0_{\rm rel} (p) \sim {\overline{\cal Z}}^0_0(p)\,/\,{\cal
D}{\overline{\cal
    Z}}^0_1(p)\;\;,
    $$
where the space ${\cal D}{\overline{\cal Z}}^0_1(p)$ of ~"gauge
states" was generated  by all $\{L_{-m}^{\rm tot}\,; m > 0\}$
operators and was identified with the radical of
    ${\cal H}_{\rm phys}(p)$
    containing all null vectors.\\
There are also null states in the anomalus case (those of discrete
series (\ref{disc_series})), but according to Vanishing Theorem
above $(\xi < 1)$ they cannot be gauged away in the sense of cohomology. \\
\\
The remaining part of this section will be devoted to the
description of the total (absolute) cohomologies of the polarized
complex. In order to reconstruct the absolute cohomology spaces
out of relative ones, one should follow the way presented in
\cite{brstn} and \cite{fbrst}. The general reasoning (which
enables one to treat the degenerate and non-degenerate cases
simultaneously) will be reported here for completeness.

First of all, it will be demonstrated that due to Vanishing
Theorem for relative cohomologies, all the absolute classes from
$H^s(p)$ are zero for $s \neq \pm \frac{1}{2}$. \\ According to
the decomposition of (\ref{relativecomplex}) any (on-mass-shell
with $p \neq 0$) element of absolute degree $s=r+\frac{1}{2}$ $(r
\in \mathbb{Z})$ can be written as:
     \begin{equation}
    \Psi^{r+\frac{1}{2}} = c_0\Phi^{r} + \Phi^{r+1}
    \;\;\;\;\;\;{\rm with}\;\;\;\;\;
    \Phi^{r},\;\Phi^{r+1} \in {\cal{P}}_{\rm rel}(p)\;\;.
 \label{z0}
 \end{equation}
Assuming that $\Psi^{r+\frac{1}{2}}$ is closed:
$D\Psi^{r+\frac{1}{2}} = 0$, one obtains the following equations for
relative components:
 \begin{equation}
 D_{\rm rel}\Phi^{r} = 0
    \;\;\;\;\;\;{\rm and}\;\;\;\;\;
    D_{\rm rel}\Phi^{r+1} + M\Phi^{r}\;\;,
 \label{z1}
 \end{equation}
where $M = \{ D\,,\,c_0\}$ is that of (\ref{relativediff}). \\
If $r \ne 0$, then according to Vanishing Theorem, the first
equation is solved by $\Phi^{r} = D_{\rm rel}\chi^{r-1}$ for some
$\chi^{r-1}$ from ${\cal{P}}^{r-1}_{\rm rel}(p)$. Inserting this
solution into second equation one obtains $D_{\rm rel}(\Phi^{r+1} +
M\chi^{r-1}) =
0$. \\
For $r \ne -1$, again due to Vanishing Theorem, one gets the
solution $\Phi^{r+1} + M\chi^{r-1} = D_{\rm rel}\xi^{r}$ for some
$\xi^{r}$ in
${\cal{P}}^{r}_{\rm rel}(p)$. \\
Making the gauge shift:
    \begin{equation}
    \Psi^{r+\frac{1}{2}} \to  {\hat \Psi}^{r+\frac{1}{2}} =
    \Psi^{r+\frac{1}{2}} + D(c_0\chi^{r-1}) = D\xi^{r}\;\;,
 \label{z2}
 \end{equation}
 one obtains the element which is cohomologous to zero. \\
 Hence all absolute cohomology spaces except of possibly
 $H^{\pm\frac{1}{2}}(p)$ are zero.

 In the second step it will be shown that the absolute classes
 $H^{\pm\frac{1}{2}}(p)$ can be reconstructed out of these of $H^{0}_{\rm
 rel}(p)$. \\
 In order to get $H^{-\frac{1}{2}}(p)$ one introduces the natural
 inclusion map on relative cocycles:
   \begin{equation}
    Z^{0}_{\rm rel}(p) \ni \Psi^{0} \to
    i_{-} (\Psi^{0}) := \Psi^{0} \in Z^{-\frac{1}{2}}(p)\;\;.
 \label{iminus}
 \end{equation}
Since $Di_{-} = i_{-}D_{\rm rel}$, the map (\ref{iminus}) is well
defined and transforms relatively exact elements into absolutely
exact ones. Hence it determines the unique mapping:
 \begin{equation}
    H^{0}_{\rm rel}(p) \ni [\Psi^{0}]_{\rm rel} \to
    i^{*}_{-} [\Psi^{0}]_{\rm rel} = [i_{-}(\Psi^{0})]_{\rm abs} \in
    H^{-\frac{1}{2}}(p)\;\;,
 \label{z3}
 \end{equation}
of the respective cohomology spaces.

The situation is a bit more complicated with $H^{\frac{1}{2}}(p)$.
It is clear that the corresponding mapping should be generated by
multiplication of relative cocycles by the ghost zero mode $c_0$. It
is however not true that simple multiplication transforms relative
cocycles into absolute ones. One has instead:
\begin{equation}
    D(c_0\Psi^{r}) = -c_0D_{\rm rel}\Psi^{r} + M\Psi^{r} \;\;.
 \label{z4}
 \end{equation}
 Nevertheless the multiplication map can be corrected in
 appropriate way. \\
 For this to be done one should observe first, that
 $M\Psi^{r}$ is always $D$-exact for $\Psi^{r}$ from $Z^{r}_{\rm
 rel}(p)$. For $r \ne 0$ and $\Psi^{r}$ in $Z^{r}_{\rm rel}(p)$, according to
 Vanishing Theorem, one has $\Psi^{r} = D_{\rm rel}\chi^{r-1}$ and
 $M\Psi^{r} = D(M\chi^{r-1})$. For $r=0$ the closed element $M\Psi^0$ is of degree 2. Hence, again
 due to Vanishing Theorem, it is cohomologically trivial: $M\Psi^0 =
 D_{\rm rel}\chi^1 = D\chi^1$ for some $\chi^1$ in ${\cal{P}}^{1}_{\rm
 rel}(p)$. \\
 Consequently, one may introduce the mapping:
 \begin{equation}
    Z^{r}_{\rm rel}(p) \ni \Psi^{r} \to
    P(M \Psi^{r}) \in {\cal{P}}^{r+1}_{\rm rel}(p)\;\;,
 \label{mapping}
 \end{equation}
which associates the primary of $M \Psi^{r}$ with any $\Psi^{r}$
from $Z^{r}_{\rm rel}$. This mapping is not uniquely determined but
it can always be chosen in such a way that $P(MD_{\rm rel}\Psi^{r})
= M\Psi^{r}$. \\
Using (\ref{mapping}) one may introduce the corrected multiplication
map:
\begin{equation}
    Z^{0}_{\rm rel}(p) \ni \Psi^{0} \to
    c(\Psi^{0}) := c_0 \Psi^{0} - P(M\Psi^{0}) \in Z^{\frac{1}{2}}(p)\;\;.
 \label{zn}
 \end{equation}
It transforms relative cocycles into absolute ones and moreover, due
to the appropriate choice of primaries, it satisfies:
\begin{equation}
    c(D_{\rm rel} \chi^{r}) = -D(c_0\chi^{r})\;\;.
 \label{zn1}
 \end{equation}
Hence one obtains well defined map of cohomology spaces:
 \begin{equation}
    H^{0}_{\rm rel}(p) \ni [\Psi^{0}]_{\rm rel} \to
    i^{*}_{+} [\Psi^{0}]_{\rm rel} = [c(\Psi^{0})]_{\rm abs} \in
    H^{\frac{1}{2}}(p)\;\;.
 \label{zn3}
 \end{equation}

 One is now in a position to prove the following
\begin{theorem}[Absolute Cohomologies]
~~~\\ The mappings: \\ $$ H^{-\frac{1}{2}}(p) \;\;
\stackrel{i_-^*}{\longleftarrow}\;\;H^{0}_{\rm
rel}(p)\;\;\stackrel{i_+^*}{\longrightarrow}\;\;H^{\frac{1}{2}}(p)\;\;\;,\;\;\;p\ne
0\;\;, $$ are the isomorphisms of the cohomology spaces.
\end{theorem}
{\it Proof} : Any cocycle $\Psi^{-\frac{1}{2}} \in
Z^{-\frac{1}{2}}(p)$ is of the form $\Psi^{-\frac{1}{2}} =
c_0\Phi^{-1} + \Phi^{0}$ with $D_{\rm rel}\Phi^{-1} = 0$ and
$D_{\rm rel}\Phi^{0} + M\Phi^{-1}= 0$. The Vanishing Theorem and
the absence of anomalus cochains of degree lower than $-1$ implies
that $\Phi^{-1} = 0$ is the unique solution. Hence
$\Psi^{-\frac{1}{2}} = \Phi^{0}$ and $i_{-}$ is onto. In order to
check injectivity one should notice that $\Psi^{-\frac{1}{2}} \sim
{\hat \Psi}^{-\frac{1}{2}}$ in the sense absolute cohomology
amounts to $\Phi^{0} - {\hat \Phi}^{0} = D_{\rm rel}\chi^{-1}$
i.e. the relative equivalence. Hence $i_{-}^{*}$ is an
isomorphism. \\ Assuming that $\Psi^{\frac{1}{2}} = c_0\Phi^{0} +
\Phi^{1}$ is a cocycle one obtains $D_{\rm rel}\Phi^{0} = 0$ and
$D_{\rm rel}\Phi^{1} + M\Phi^{0} = 0$. From the first equation it
follows that $M\Phi^{0}$ is closed and due to the considerations
above one can take its primary to solve for $\Phi^{1}$: $\Phi^{1}
= -P(M\Phi^{0})$. Hence $\Psi^{\frac{1}{2}} = c_0\Phi^{0} -
P(M\Phi^{0}) = c(\Phi^{0})$. Consequently the map $i_{+}^{*}$ is
onto. Assume that $c(\Phi^{0}) \sim c({\hat \Phi}^{0})$ in the
absolute space. This amounts to $c_0(\Phi^{0} - {\hat \Phi}^{0}) -
P(M\Phi^{0}) + P(M{\hat \Phi}^{0}) = - c_0D_{\rm rel}\chi^0 +
M\chi^0 + D_{\rm rel}\chi^1$. Hence $\Phi^{0} - {\hat \Phi}^{0} =
-D_{\rm rel}\chi^0$ and $\chi^1 = -P(M\chi^0)$. This proves
injectivity of $i_+^*$.$\;\;\Box$ \\
\\
The general results obtained above together with 2) of Vanishing
Theorem for relative cohomologies give  the tractable
representation of the absolute cohomology classes.\\
    In the non-degenerate case one has\footnote{By slight abuse of notation, the space
    ${\cal H}_{\rm phys}$ of (\ref{phys}) is identified with its image in the complex
    ${\cal H}_{\rm phys}\otimes \omega$, where $\omega$ denotes the ghost vacuum. }
   $$
   H^{-\frac{1}{2}}(p) \simeq {\cal H}_{\rm phys}(p)\;\;\;\;{\rm
   and}\;\;\;\;H^{\frac{1}{2}}(p) \simeq c_0{\cal H}_{\rm phys}(p)
   \;,$$
while for degenerate anomaly one obtains the quotients:
    $$
 H^{-\frac{1}{2}}(p) \simeq {\cal H}_{\rm phys}(p)/\sim \;\;\;\;{\rm
   and}\;\;\;\;H^{\frac{1}{2}}(p) \simeq c_0{\cal H}_{\rm
   phys}(p)/\sim
   \;.
    $$
The equivalence relation above is defined by: $\Psi \sim \Psi'$ if
and only if  $\Psi - \Psi' = L_{-1} \Phi$ for some $\Phi$ from
${\cal H}_{\rm phys}$.\\
\\
 All the
results of this chapter were obtained under essential assumption
that the momentum underlying the polarized anomalus complex was
non-zero. \\
\\
There is a series of solutions of the mass-shell equation
(\ref{massshell}) admitting $p=0$. They are all located in the
discrete series (\ref{disc_series}) of unitary models. The
numerical calculations show that there are perfect vacuum states
in several spacetime dimensions at level $N=0$, and zero momentum
states at first excited level $N=1$ in $d=25$. For example, in
$d=4$, the perfect vacuum states exist for $m=24$, $r=s=20$ and
$m=242$, $r=s=198$. The zero momentum states at level $N=1$ are
admissible for $d=25$ and $m=24$, $r=s=10$ or $m=242$, $r=s=99$.
The numerical calculations performed up to $m=900$ indicate that
one finds less and less solutions with $m$ growing.

The absolute cohomologies for $p = 0$ case can be calculated
directly. One finds immediately:
 \begin{equation}
    H^{-\frac{1}{2}}(0) = \mathbb{C}\Omega(0)\;\;\;,\;\;\;H^{\frac{1}{2}}(0) =
    \mathbb{C}c_0\Omega(0)\;\;\;{\rm for}\;\; N=0\;\;,
 \label{cohozero1}
 \end{equation}
and
$$
    H^{-\frac{1}{2}}(0) = {\cal H}_{\rm phys}^{1}(0),\;\;\;H^{\frac{1}{2}}(0) =
    c_0{\cal H}_{\rm phys}^{1}(0)\oplus \mathbb{C}c_{-1}\Omega(0),\;\;\;H^{\frac{3}{2}}(0) =
    \mathbb{C}c_0c_{-1}\Omega(0)\;\;,
    \nonumber
$$ for $N=1$, where the space ${\cal H}_{\rm phys}^{1}(0)$ above
is generated by level one matter modes $a_{-1}^\mu$, $u_{-1}$;
$\mu = 0 , \ldots, d - 1$ out of the perfect matter/ghost vacuum
state $\Omega(0)$.

One should notice the lack of duality at $N=1$. The state
$c_{-1}\Omega(0)$ cannot be eliminated because it's primary
$c_0b_{-1}\Omega(0)$ does not belong to polarized anomalus
complex. For the same reason the dual partner $b_{-1}\Omega(0)$ of
$H^{\frac{3}{2}}(0)$ is missing in the kernel of $J_{\xi}^{(+)}$.
\\
\\
The above considerations exhaust and finish the analysis of polarized anomalus
cohomologies for non-critical massive string theories.

\section{Final remarks}
The cohomologies of the polarized anomalus complex for
non-critical massive strings models were investigated in this
paper. The proposed approach, although restricted to string
models, seems to be an universal tool to describe the constrained
systems with anomalies and possibly to develop a theory of their
interactions.

The results of this paper indicate that the polarized anomalus
complexes introduced here should give the resolutions of the
spaces of physical states for the wide class of models with
anomalies. It has to be stressed that the presented method does
not distinguish the cases of whether the anomalies are of
classical or quantum origin or both together.

The authors did not calculate the cohomologies of the full
anomalus complex. This problem is left open. It would be
interesting to determine them and to find their geometrical
meaning and physical interpretation.

There is also an interesting problem related with the non-degenerate
pairing on the anomalus complex and its polarized subcomplex. It
appears that it can be introduced by the use of the anomaly itself.
The anomaly defines the pseudo-Kaehler structure on ${\cal A}$ and
${\cal P}$. The corresponding Kaehler-Laplace operators can be used
to define the harmonic representatives of the non-vanishing classes.
The work on this subject is advanced and the results will be
presented in the forthcoming paper.

The authors intend to investigate the properties of the proposed
approach in much wider context. The work on finite dimensional
systems with anomalies (of the classical origin obviously) is in
progress. The authors aim to investigate the field-theoretical
models (with anomalies of both kinds - quantum and classical) in
near future.

\bigskip\bigskip
\noindent {\bf Acknowledgements}\\ One of us (M.D.) would like to
thank Institute of Theoretical Physics UwB for warm hospitality
during his stays in Bia{\l}ystok and  his friend Grzegorz Stachowicz
for his great heart. 

\noindent

\begin{thebibliography}{99}
\bibitem{hasjas} Z.Hasiewicz, Z.Jask\'{o}lski,
Nucl.Phys. B464 (1996) 85
\bibitem{bdhp} L.Brink, P. di Vecchia and P.Howe, Nucl.Phys. B118
(1977) 76\\
A.M.Polyakov, Phys. Lett. B103 (1981) 207
\bibitem{gb} S.Gupta, Proc. Roy. Soc. A63 (1950) 681\\
K.Bleuler, Helv. Phys. Acta 23 (1950) 567
\bibitem{noghost} D.Friedan, Z.Qiu, S.Shenker, Phys. Rev. Lett. 52
(1984) 1575\\
D.Friedan, Z.Qiu, S.Shenker, Commun. Math. Phys 107 (1986) 535\\
P.Goddard, A.Kent, D.Olive, Phys. Lett. B152 (1985) 88
\bibitem{spectrum} M.Daszkiewicz, Z.Hasiewicz, Z.Jask\'{o}lski,
Nucl.Phys. B514 (1998) 437 \\
M.Daszkiewicz, Z.Jask\'{o}lski, Phys. Lett. B484 (2000) 347
\bibitem{brstinteractions} E. Witten, Nucl.Phys. B268 (1986) 253\\
D.Friedan, E.Martinec, S.Shenker, Nucl.Phys. B271 (1986) 93\\
H.Hata, K.Itoh, T.Kugo, H.Kunimoto, K.Ogawa, Phys.Rev. D34 (1986)
2360\\
T.Banks, M.E.Peskin, Nucl.Phys. B264 (1986) 513
\bibitem{brstn} Z.Hasiewicz, J.Phys. A33 (2000) 7773
\bibitem{fradkin} E.S.Fradkin, G.A.Vilkovisky, Phys. Lett. B55 (1975) 224\\
I.A.Batalin, G.A.Vilkovisky, Phys. Lett. B69 (1977) 309\\
E.S.Fradkin, T.E.Fradkina, Phys. Lett. B72 (1978) 343\\
B.Lian, G.J. Zuckermann, Commun.Math.Phys. 125 (1989) 301
\bibitem{GB}S.Aoyama, J.Kowalski-Glikman, J.Lukierski, J.W. van
Holten, Phys.Lett. B216 (1989) 133\\
Z.Hasiewicz, J.Kowalski-Glikman, J.Lukierski, J.W. van
Holten, Phys.Lett. B217 (1989) 95\\
Z.Hasiewicz, J.Kowalski-Glikman, J.Lukierski, J.W. van Holten,
J.Math.Phys. 32 (1991) 2358\\
J.Kowalski-Glikman, Annals Phys. 232 (1994) 1
\bibitem{Mordechai} M.Spigelglass, Nucl.Phys. B283 (1987) 205
\bibitem{difrancesco} P.Di Francesco, P.Mathieu {\it Conformal field
theory} Springer Verlag, New York (1997)
\bibitem{libermann} P. Libermann, Ch-M.Marle {\it Sympletic Geometry
and Analytical Mechanics} D.Reidel, Dordrecht, Boston, Lancaster,
Tokio (1987)\\
V.Guillemiu, S.Sternberg {\it Sympletic techniques in Physics}
Cambridge University Press, Cambridge (1984)
\bibitem{GHV} W.Greub, S.Halperin, R.Vanstone {\it
Conections, Curvature, and Cohomology} vol. III Academic Press, New
York (1976)
\bibitem{hwang} S.Hwang, Phys.Rev. D28 (1983) 2614
\bibitem{Wells} R.O.Wells, Jr. {\it Differential Analysis on Complex manifolds}
Prentice Hall, Englewood Cliffs, N.J. (1973)\\ S.-s.Chern {\it
Complex Manifolds Without Potential Theory} Springer Verlag, New
York (1979)
\bibitem{bourbaki} BOURBAKI {\it Groupes et Algebres de Lie}
Chapitre VII-VIII, Herman Paris (1975)
\bibitem{BMP} P.Bouwknegt, J.McCarthy, K.Pilch, Commun.Math.Phys
145  (1992) 54
\bibitem{BT} R.Bott, L.W.Tu {\it Differential forms in algebraic
topology} Springer Verlag, New York 1982
\bibitem{fbrst} Z.Hasiewicz, J.Phys. A34 (2001) 1861

\end{thebibliography}
\end{document}